\documentclass[journal]{IEEEtran}
\makeatletter\if@twocolumn\PassOptionsToPackage{switch}{lineno}\else\fi\makeatother

      \makeatletter
\usepackage{wrapfig}
\newcounter{aubio}

\long\def\bioItem{%
\@ifnextchar[{\@bioItem}{\@@bioItem}}

\long\def\@bioItem[#1]#2#3{
 \stepcounter{aubio}
 \expandafter\gdef\csname authorImage\theaubio\endcsname{#1}
 \expandafter\gdef\csname authorName\theaubio\endcsname{#2}
 \expandafter\gdef\csname authorDetails\theaubio\endcsname{#3}
}

\long\def\@@bioItem#1#2{
 \stepcounter{aubio}
 \expandafter\gdef\csname authorName\theaubio\endcsname{#1}
 \expandafter\gdef\csname authorDetails\theaubio\endcsname{#2}
}

\newcommand{\checkheight}[1]{%
  \par \penalty-100\begingroup%
  \setbox8=\hbox{#1}%
  \setlength{\dimen@}{\ht8}%
  \dimen@ii\pagegoal \advance\dimen@ii-\pagetotal
  \ifdim \dimen@>\dimen@ii
    \break
  \fi\endgroup}

\def\printBio{%
  \@tempcnta=0
   \loop
     \advance \@tempcnta by 1
     \def\aubioCnt{\the\@tempcnta}
     \setlength{\intextsep}{0pt}%
     \setlength{\columnsep}{10pt}%
     \newbox\boxa%
     \setbox\boxa\vbox{\csname authorDetails\aubioCnt\endcsname}
     \expandafter\ifx\csname authorImage\aubioCnt\endcsname\relax%
      \else%
       \checkheight{\includegraphics[height=1.25in,width=1in,keepaspectratio]{\csname authorImage\aubioCnt\endcsname}}
        \begin{wrapfigure}{l}{25mm}
         \includegraphics[height=1.25in,width=1in,keepaspectratio]{\csname authorImage\aubioCnt\endcsname}
        \end{wrapfigure}\par
      \fi
     {\parindent0pt\textbf{\csname authorName\aubioCnt\endcsname}\csname authorDetails\aubioCnt\endcsname \par\bigskip%
     \expandafter\ifx\csname authorImage\aubioCnt\endcsname\relax\else%
      \ifdim\the\ht\boxa < 90pt\vskip\dimexpr(90pt -\the\ht\boxa-1pc)\fi%
     \fi}
      \ifnum\@tempcnta < \theaubio
   \repeat
   }

\makeatother

\usepackage{graphicx}
\usepackage[T1]{fontenc}
\ifCLASSINFOpdf
\else
\fi
\usepackage{amsmath}
\usepackage[cmintegrals]{newtxmath}
\usepackage{url,multirow,morefloats,floatflt,cancel,tfrupee}
\makeatletter

\AtBeginDocument{\@ifpackageloaded{textcomp}{}{\usepackage{textcomp}}}
\makeatother
\usepackage{colortbl}
\usepackage{xcolor}
\usepackage{pifont}
\usepackage[nointegrals]{wasysym}
\makeatletter

\def\mcWidth#1{\csname TY@F#1\endcsname+\tabcolsep}

\def\cAlignHack{\rightskip\@flushglue\leftskip\@flushglue\parindent\z@\parfillskip\z@skip}
\def\rAlignHack{\rightskip\z@skip\leftskip\@flushglue \parindent\z@\parfillskip\z@skip}

\@ifundefined{etal}{}{}

\usepackage{ifxetex}
\ifxetex\else\if@twocolumn\@ifpackageloaded{stfloats}{}{\usepackage{dblfloatfix}}\fi\fi

\AtBeginDocument{
\expandafter\ifx\csname eqalign\endcsname\relax
\def\eqalign#1{\null\vcenter{\def\\{\cr}\openup\jot\m@th
  \ialign{\strut$\displaystyle{##}$\hfil&$\displaystyle{{}##}$\hfil
      \crcr#1\crcr}}\,}
\fi
}

\AtBeginDocument{%
  \@ifpackageloaded{endfloat}%
   {\renewcommand\efloat@iwrite[1]{\immediate\expandafter\protected@write\csname efloat@post#1\endcsname{}}}{\newif\ifefloat@tables}%
}%

\def\BreakURLText#1{\@tfor\brk@tempa:=#1\do{\brk@tempa\hskip0pt}}
\let\lt=<
\let\gt=>
\def\processVert{\ifmmode|\else\textbar\fi}

\@ifundefined{subparagraph}{
\def\subparagraph{\@startsection{paragraph}{5}{2\parindent}{0ex plus 0.1ex minus 0.1ex}%
{0ex}{\normalfont\small\itshape}}%
}{}

\newcommand\role[1]{\unskip}
\newcommand\aucollab[1]{\unskip}
  
\@ifundefined{tsGraphicsScaleX}{\gdef\tsGraphicsScaleX{1}}{}
\@ifundefined{tsGraphicsScaleY}{\gdef\tsGraphicsScaleY{.9}}{}
\def\checkGraphicsWidth{\ifdim\Gin@nat@width>\linewidth
	\tsGraphicsScaleX\linewidth\else\Gin@nat@width\fi}

\def\checkGraphicsHeight{\ifdim\Gin@nat@height>.9\textheight
	\tsGraphicsScaleY\textheight\else\Gin@nat@height\fi}

\def\fixFloatSize#1{}
\let\ts@includegraphics\includegraphics

\def\inlinegraphic[#1]#2{{\edef\@tempa{#1}\edef\baseline@shift{\ifx\@tempa\@empty0\else#1\fi}\edef\tempZ{\the\numexpr(\numexpr(\baseline@shift*\f@size/100))}\protect\raisebox{\tempZ pt}{\ts@includegraphics{#2}}}}

\AtBeginDocument{\def\includegraphics{\@ifnextchar[{\ts@includegraphics}{\ts@includegraphics[width=\checkGraphicsWidth,height=\checkGraphicsHeight,keepaspectratio]}}}

\DeclareMathAlphabet{\mathpzc}{OT1}{pzc}{m}{it}

\def\URL#1#2{\@ifundefined{href}{#2}{\href{#1}{#2}}}

\def\UrlOrds{\do\*\do\-\do\~\do\'\do\"\do\-}%
\g@addto@macro{\UrlBreaks}{\UrlOrds}

\edef\fntEncoding{\f@encoding}

\makeatother

\newif\ifmultipleabstract\multipleabstractfalse%
\usepackage{tabulary}
\makeatletter
\AtBeginDocument{\@ifpackageloaded{longtable}{%
\def\LT@makecaption#1#2#3{%
  \LT@mcol\LT@cols c{\hbox to\z@{\hss\parbox[t]\LTcapwidth{%
    \sbox\@tempboxa{#1{#2: } #3}%
    \ifdim\wd\@tempboxa>\hsize
      #1{#2: }\textsc{#3}%
    \else
      \hbox to\hsize{\hfil\box\@tempboxa\hfil}%
    \fi
    \endgraf\vskip\baselineskip}%
  \hss}}}
}{}}
\makeatother

   \makeatletter
  \def\fig@textbf{\textbf}
   \AtBeginDocument{\renewcommand\floatc@plain[2]{\setbox\@tempboxa\hbox{{\footnotesize#1.}\footnotesize\hskip.5em#2}%
    \ifdim\wd\@tempboxa>\hsize {\fig@textbf{\footnotesize#1.}}\footnotesize\hskip.5em#2\par
        \else\hbox to\hsize{\hfil\box\@tempboxa\hfil}\fi}}
    \makeatother
  
\usepackage{float}

\begin{document}

%

        \title{A New Update Rule of ADFF-RLS \& EKF based Joint-estimation Filters for Real-time SOC/SOH Identification}
      
\author{Kwangrae~Kim,
        Minho~Kim,
        Suwon~Kang,
        Jungwook~Yu,
        Jungsoo~Kim,
        Huiyong~Chun, and 
        Soohee~Han\thanks{Kwangrae~Kim (e-mail: caritas@postech.ac.kr)}\thanks{
        Minho~Kim (e-mail: minho1st@postech.ac.kr)}\thanks{
        Jungwook~Yu (e-mail: jyu14@postech.ac.kr)}\thanks{
        Jungsoo~Kim (e-mail: jungsoo.kim@postech.ac.kr)}\thanks{
        Huiyong~Chun (e-mail: ymir1994@postech.ac.kr)}\thanks{Corresponding author: 
        Soohee~Han (e-mail: sooheehan@postech.ac.kr)}}

\maketitle 

\begin{abstract}
In order to accurately estimate the SOC and SOH of a lithium-ion battery used in an electric vehicle (EV), we propose an Adaptive Diagonal Forgetting Factor Recursive Least Square (ADFF-RLS) for accurate battery parameter estimation. ADFF-RLS includes two new proposals in the existing DFF-RLS; The first is an excitation tag that changes the behavior of the DFF-RLS and the EKF according to the dynamics of the input data. The second is auto-tuning that automatically finds the optimal value of RLS forgetting factor based on condition number (CN). Based on this, we proposed a joint estimation algorithm of ADFF-RLS and Extended Kalman Filter (EKF). To verify the accuracy of the proposed algorithm, we used experimental data of hybrid pattern battery cells mixed with dynamic and static patterns. In addition, we added a current measurement error that occurs when measuring at EV, and realized data that is closer to actual environment. This data was applied to two conventional estimation algorithms (Coulomb counting, Single EKF), two joint estimation algorithms (RLS \& EKF, DFF-RLS \& EKF) and ADFF-RLS \& EKF. As a result, the proposed algorithm showed higher SOC and SOH estimation accuracy in various driving patterns and actual EV driving environment than previous studies.
\end{abstract}
    

%
\IEEEpeerreviewmaketitle

\section{Introduction}
In response to the demand for reducing pollution, Electric vehicles (EVs) are becoming increasingly common throughout the world. In general, EVs use Li-ion batteries with a variety of advantages over other batteries as energy storage devices.\unskip~\cite{Cite1} At this time, a battery management system (BMS) that controls the battery is important for increasing the life of the battery and maintaining safety.

 To control the battery efficiently, BMS needs to know the current internal state of the battery, the two most important states being state-of-charge (SOC) indicating current charge of battery and state-of-charge (SOH). The SOC basically represents the distance the vehicle can travel, and SOH represents the current and remaining service life of the EV battery. Therefore, real-time estimation of SOC and SOH is important for improving the efficiency of the driving cost, driving convenience and economy of the vehicle.

As a representative method for estimating SOC and SOH, there is a model based method in which a system model in which the input of the battery is used as the current and the output is the voltage is established and the internal state is estimated based on the system model. In general, the commercial BMS of the EV is an Equivalent Circuit Model (ECM), which is easy to apply and has low computational complexity.\unskip~\cite{342681:7604108}

Various SOC/SOH estimation algorithms based on ECM have been studied. Among them, filter based method and observer based method are mainly used in commercial BMS. \unskip~\cite{342681:7589621,342681:7589623,342681:7589626,342681:7589631,342681:7589629}. However, when the existing algorithms are used solely to estimate the state of the EV, it is generally possible to estimate only the SOC alone and does not adequately cope with changes in SOH and characteristic of the battery which changes in real time. Therefore, algorithms that can estimate the SOH by simultaneously improving the real - time estimation performance of the SOC by separating the battery parameter estimation method and the state estimation method have been proposed. This method is called joint-estimation or dual-estimation.\unskip~\cite{342681:7589622,342681:7589630}In recent papers, joint-estimation is commonly used, so in this paper we will refer to it as follows. \unskip~\cite{342681:7589630,342681:7589628,342681:7589624}

The most common structure of joint estimation is to combine two different algorithms to estimate parameters and states, respectively. Furthermore, recently, the most studied combinations are using RLS for parameter estimation and KF for state estimation.\unskip~\cite{342681:15364123}  The advantage of this combination is that it has high battery parameter estimation performance and high SOC/SOH estimation performance while having a level of computation applicable to BMS. These methods commonly use conventional RLS to apply a single forgetting factor value (FV) to RLS. When these methods are applied to the actual driving pattern of the EV, the RLS parameter estimation performance decreases significantly in a certain period.

To solve this problem, a new SOC \& SOH estimation algorithm is proposed based on the following new techniques. First, we propose a joint estimation algorithm of the Extended Kalman Filter and the Diagonal Forgetting Factor RLS (DFF-RLS) that estimates the SOC and SOH based on the real-time parameter estimation in the BMS of the EV. Second, we propose the auto-tuning method of DFF-RLS using condition number to improve 'wind-up problem' which is weakness of existing RLS-based technique and to estimate more accurate parameter. Third, we proposed an auto-tuning method to apply the optimal forgetting factor of DFF-RLS in real time using condition number.

This paper is organized as follows. Section 2 describes the equivalent circuit model (ECM) and the conventional algorithms (RLS, DMFF-RLS, EKF). Section 3 describes the proposed adaptive multiple forgetting factor recursive least square (ADFF-RLS). Section 4 introduces Experimental result, which verifies the performance of the proposed method using real data. Finally, Section 5 describes the overall summary and future work.
    
\section{Joint estimation using RLS \& EKF}

\subsection{Equivalent Circuit Model}There are various models representing the battery. Among them, the Equivalent Circuit Model (ECM) model, which has low complexity and high model reliability and accuracy, is easy to apply to actual applications. In general, it is known that the 1RC model can express the dynamics of a battery relatively accurately with less computational burden for a Li-NMC battery. \unskip~\cite{342681:7604108,342681:13254944} For this reason, it is widely used in BMS of commercial EV, and this paper also used 1RC model.

\bgroup
\fixFloatSize{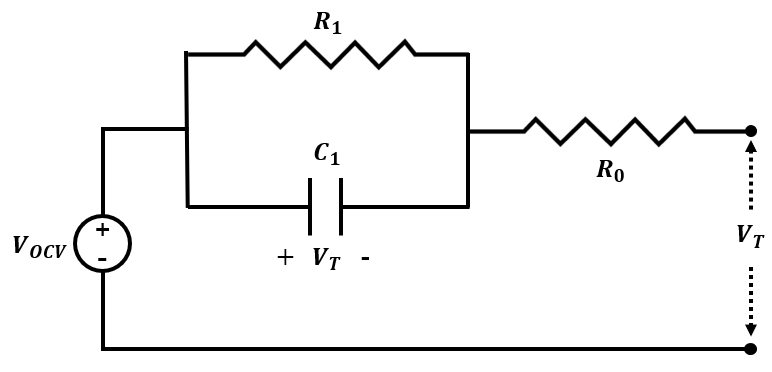}
\begin{figure}[!htbp]
\centering \makeatletter\IfFileExists{images/A5.png}{\includegraphics{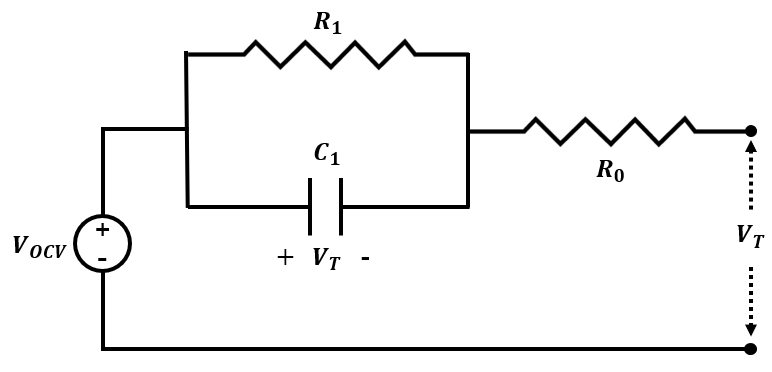}}{}
\makeatother 
\caption{{1RC Equivalent circuit model}}
\label{figure-9b4a0e28b62377f918687c7fbd786daf}
\end{figure}
\egroup
 The 1RC model consists of one voltage source, one series resistor, and one series RC branch. Where $V_L $is the terminal voltage of battery, $V_{OCV} $ is the open circuit voltage, $I_L $ is the terminal current, The branch consist of $R_1,\;C_1 $ is equivalent polarization resistance and capacitance represent slow response, $R_0 $ is equivalent serial ohmic resistance which represents instantaneous response.

\subsection{Joint estimation using DFF-RLS \& EKF}Because the characteristics of the battery change in real time, it is necessary to update the model parameters. Recently, `Joint estimation' which combines parameter update algorithm and state estimation algorithm has been studied extensively. By using this method, we can obtain high state estimation performance while compensating for the disadvantages of each algorithm.\unskip~\cite{342681:8204099,342681:8204101}

\bgroup
\fixFloatSize{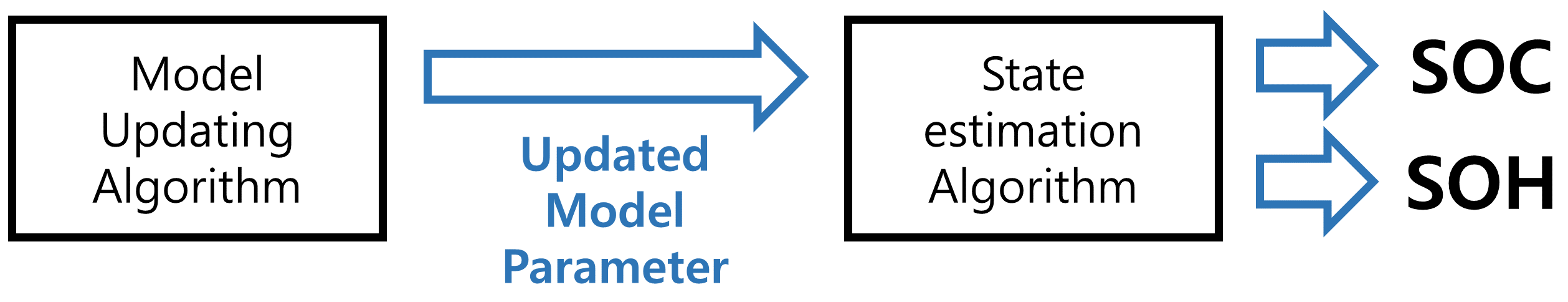}
\begin{figure}[!htbp]
\centering \makeatletter\IfFileExists{images/A1.png}{\includegraphics{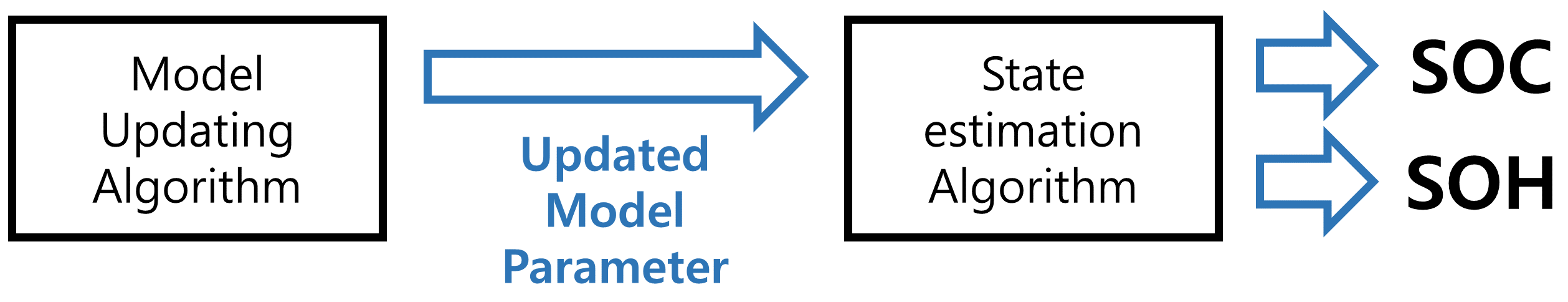}}{}
\makeatother 
\caption{{Structure of Joint estimation}}
\label{figure-f1730111690e0b448395db7d320c479f}
\end{figure}
\egroup
In general, EKF and RLS are the most commonly used joint-estimation algorithms for SOC/SOH estimation. EKF is most common algorithm for estimating the state of a system, and RLS is widely used for estimating and tracking time-varying parameters.\unskip~\cite{342681:8281700,342681:8281701}

In this paper, Joint-estimation type SOC \& SOH real-time estimation algorithm combining DFF-RLS and EKF is proposed. DFF-RLS is an algorithm that can adapt forgetting to the characteristics of parameters that estimate FV as n-dim diagonal matrix. It is highly effective when the sensitivity of parameters such as battery is considerably different.

\subsection{Extended Kalman Filter (EKF)} Extended kalman filter (EKF) is optimal state observer for nonlinear system which estimates true state recursively. EKF estimates system state from system model and measurement.\unskip~\cite{342681:13001505} At this time, EKF is linearized by finding jacobian of the nonlinear model equation f(x). This method has the advantage of easily applying KF to nonlinear models.

This algorithm performs a `primary estimate' to estimate the state of the past data, and then performs a `posteriori-estimate' to correct the value obtained from the primary using the sensor value. Fig. 2.5 conceptually illustrates the operation of the EKF when system state equation.Equations~(\ref{dfg-6a65042c081c}) and~(\ref{dfg-fb60485ac8ac})
\let\saveeqnno\theequation
\let\savefrac\frac
\def\dispfrac{\displaystyle\savefrac}
\begin{eqnarray}
\let\frac\dispfrac
\gdef\theequation{2.1}
\let\theHequation\theequation
\label{dfg-6a65042c081c}
\begin{array}{@{}l}X(k+1)\quad =\quad AX(k)+Bu(k)+Q\end{array}
\end{eqnarray}
\global\let\theequation\saveeqnno
\addtocounter{equation}{-1}\ignorespaces 

\let\saveeqnno\theequation
\let\savefrac\frac
\def\dispfrac{\displaystyle\savefrac}
\begin{eqnarray}
\let\frac\dispfrac
\gdef\theequation{2.2}
\let\theHequation\theequation
\label{dfg-fb60485ac8ac}
\begin{array}{@{}l}Y(k+1)\quad =\quad HX(k+1)+R\end{array}
\end{eqnarray}
\global\let\theequation\saveeqnno
\addtocounter{equation}{-1}\ignorespaces 

\bgroup
\fixFloatSize{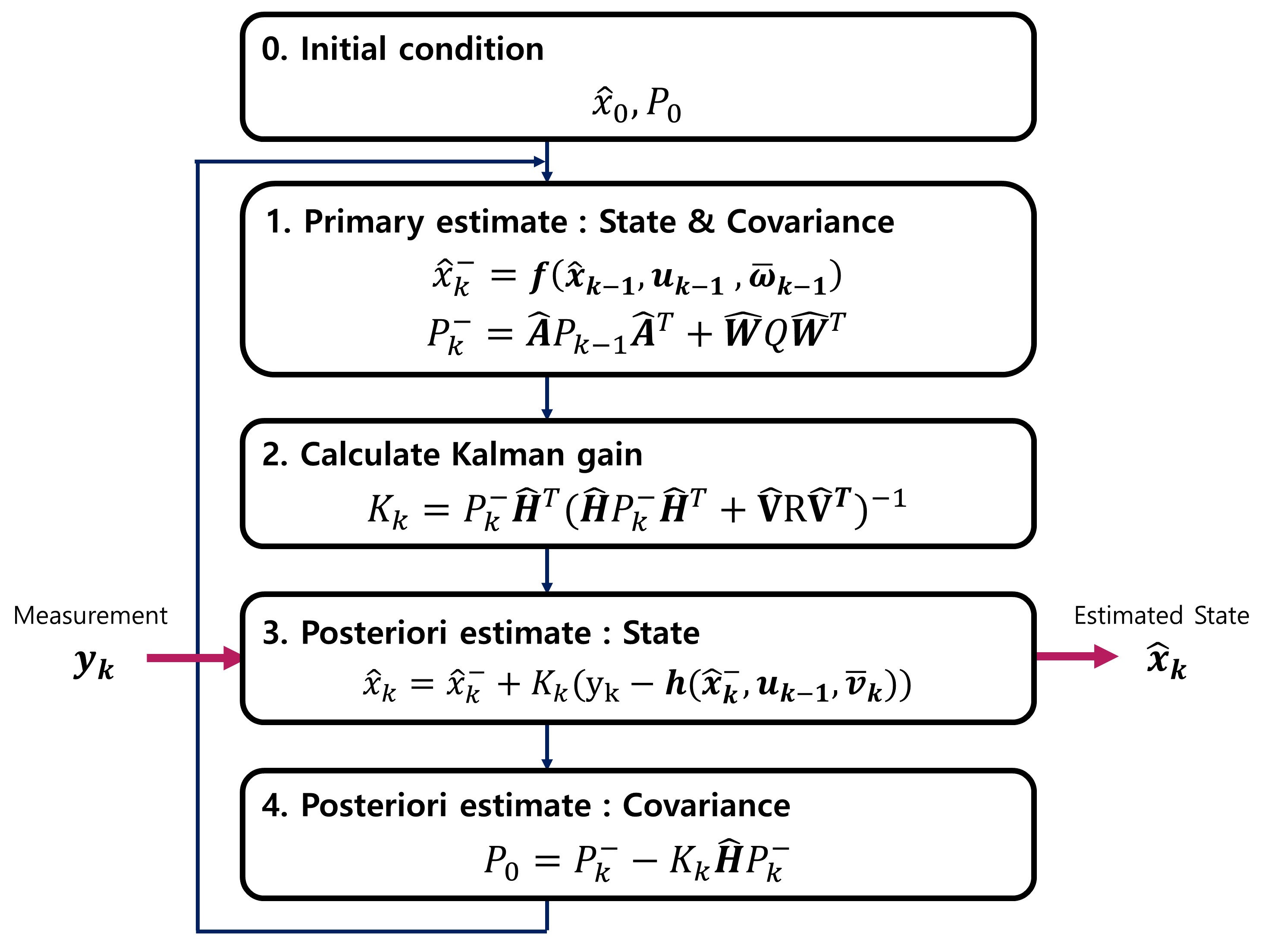}
\begin{figure}[!htbp]
\centering \makeatletter\IfFileExists{images/A2.jpg}{\includegraphics{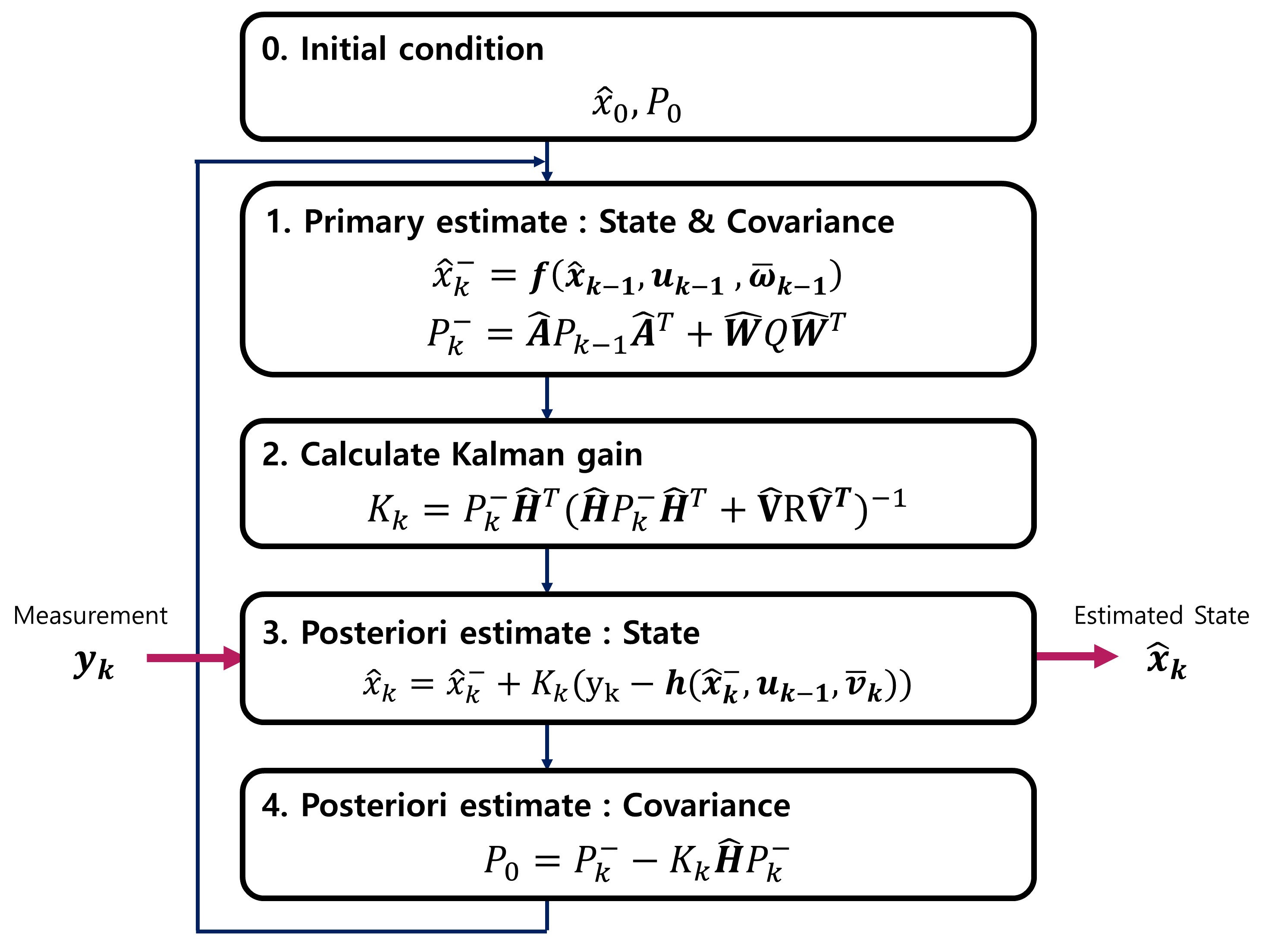}}{}
\makeatother 
\caption{{Diagram of overall EKF algorithm}}
\label{f-d0a38c6e73b6}
\end{figure}
\egroup
Where ${\widehat x}_{k-1},\;{\widehat x}_k $ are primary estimated state and posteriori estimated state, A, B, H are system model, Q, R are noise covariance, P, K are updating covariance and Kalman gain. 
\let\saveeqnno\theequation
\let\savefrac\frac
\def\dispfrac{\displaystyle\savefrac}
\begin{eqnarray}
\let\frac\dispfrac
\gdef\theequation{2.3}
\let\theHequation\theequation
\label{dfg-8efc4291f9bb}
\begin{array}{@{}l}\begin{bmatrix}SOC_{k+1}\\I_{k+1}\end{bmatrix}=\begin{bmatrix}1&0\\0&A_{RC}\end{bmatrix}\begin{bmatrix}SOC_k\\I_k\end{bmatrix}+\begin{bmatrix}\frac{-\eta\lbrack k\rbrack\Delta t}Q&0\\B_{RC}&0\end{bmatrix}\begin{bmatrix}I_k\\0\end{bmatrix}\end{array}
\end{eqnarray}
\global\let\theequation\saveeqnno
\addtocounter{equation}{-1}\ignorespaces 
 We use the 1RC model which is widely used in real BMS's in the equation of state equation of EKF. Generally, it is known that the model accuracy increases as the number of RC branches increases in the nRC model. However, there is a disadvantage that the computational burden increases due to the increase of the number of parameters to be estimated and the nonlinearity.\unskip~\cite{342681:7604108,342681:13254944}

\subsection{Diagonal Forgetting factor Recursive Least Square (DFF-RLS)}When parameters are estimated using a conventional RLS, there is a phenomenon in which the estimation accuracy is degraded in an area where data excitation is insufficient. This is called the `estimator wind-up' or 'covariance wind-up'. One reason for this problem in conventional RLS is that it applies the same covariance matrix and FV to battery parameters with different changes. When applying the same error covariance and FV for parameters that show different changes, errors due to all parameters are lumped into one single scalar term. As a result, drift occurs in one parameter estimation process, and the influence of the drift affects all parameter estimates, thereby reducing the overall estimation performance. It is also difficult to design a FV optimized for the variation characteristics of each parameter. To solve this problem, it is necessary to apply covariance matrix and FV for each parameter.

 To solve this problem, we used diagonal forgetting factor -RLS (DFF-RLS) as a parameter estimation algorithm in this paper. This algorithm can apply forgetting to the characteristics of the parameters estimating the FV as an n-dim diagonal matrix. It is highly effective when the sensitivity of parameters such as battery is considerably different.

In this algorithm, when given data matrix $\phi $  and diagonal forgetting factor matrix $\Lambda  $
\let\saveeqnno\theequation
\let\savefrac\frac
\def\dispfrac{\displaystyle\savefrac}
\begin{eqnarray}
\let\frac\dispfrac
\gdef\theequation{2.2}
\let\theHequation\theequation
\label{dfg-11a7752d85fb}
\begin{array}{@{}l}\phi(t)=\begin{bmatrix}\begin{array}{c}1\\v(t-1)\\I(t)\\I(t-1)\end{array}\end{bmatrix}\end{array}
\end{eqnarray}
\global\let\theequation\saveeqnno
\addtocounter{equation}{-1}\ignorespaces 

\let\saveeqnno\theequation
\let\savefrac\frac
\def\dispfrac{\displaystyle\savefrac}
\begin{eqnarray}
\let\frac\dispfrac
\gdef\theequation{2.3}
\let\theHequation\theequation
\label{dfg-fc32e713bd8a}
\begin{array}{@{}l}\Lambda=\begin{bmatrix}\begin{array}{ccc}\lambda_1&0&0\\0&\ddots&0\\0&0&\lambda_N\end{array}\end{bmatrix}\end{array}
\end{eqnarray}
\global\let\theequation\saveeqnno
\addtocounter{equation}{-1}\ignorespaces 
First, we need to initialize parameter $\theta(0) $  and error covariance $P_0 $ .

After that, DFF-RLS calculates updating gain(K) : 
\let\saveeqnno\theequation
\let\savefrac\frac
\def\dispfrac{\displaystyle\savefrac}
\begin{eqnarray}
\let\frac\dispfrac
\gdef\theequation{2.4}
\let\theHequation\theequation
\label{dfg-06d0}
\begin{array}{@{}l}K\left(t\right)=\frac{P\left(t-1\right)\mathrm\Lambda^{-1}\varphi(t)}{I+\varphi\left(t\right)^{T}P(t-1)\mathrm\Lambda^{-1}\varphi\left(t\right)}\end{array}
\end{eqnarray}
\global\let\theequation\saveeqnno
\addtocounter{equation}{-1}\ignorespaces 
and estimation error ($\alpha $) : 
\let\saveeqnno\theequation
\let\savefrac\frac
\def\dispfrac{\displaystyle\savefrac}
\begin{eqnarray}
\let\frac\dispfrac
\gdef\theequation{2.5}
\let\theHequation\theequation
\label{dfg-637d}
\begin{array}{@{}l}\alpha\left(t\right)=y\left(t\right)-\varphi\left(t\right)^{T}\theta(t-1)\end{array}
\end{eqnarray}
\global\let\theequation\saveeqnno
\addtocounter{equation}{-1}\ignorespaces 
for each time and updates them. Using this values, we can estimate the model parareter ($\theta $) and error covariance (P)
\let\saveeqnno\theequation
\let\savefrac\frac
\def\dispfrac{\displaystyle\savefrac}
\begin{eqnarray}
\let\frac\dispfrac
\gdef\theequation{2.6}
\let\theHequation\theequation
\label{dfg-7957}
\begin{array}{@{}l}\theta\left(t\right)=\theta\left(t-1\right)+K\left(t\right)\alpha(t)\end{array}
\end{eqnarray}
\global\let\theequation\saveeqnno
\addtocounter{equation}{-1}\ignorespaces 
where $\theta $  is estimated parameter, $\varphi $ is sensor input matrix composed of voltage and current. 
    
\section{Adaptive Diagonal Forgetting Factor Recursive Least Square (ADFF-RLS)}

\subsection{Real EV's driving pattern}The driving pattern of actual EVs is a hybrid profile with a mixture of static profile (rest and CC-CV charging) and dynamic profile (driving). Therefore, the algorithm to be mounted on the BMS of the actual EV should show high estimation performance in the hybrid profile. However, most of the existing studies have only verified the performance using only the dynamic (driving) profile represented by UDDS\unskip~\cite{342681:13001590} and NEDC\unskip~\cite{342681:13255303}. Therefore, when estimating parameters in real time in a hybrid profile, the estimation accuracy is significantly reduced. In addition, normal estimation is not performed in the static profile section. Therefore, the algorithms currently being studied may show lower performance than the simulation results when applied to the BMS of the actual EV. In general, even if the algorithm has high estimation accuracy in the dynamic profile, when the parameter is estimated in real time in the hybrid profile, the estimation accuracy decreases considerably. In particular, normal estimation is not made in the static profile section in the middle of the entire profile. In the case of RLS, covariance wind-up phenomenon occurs in the static profile section, and normal parameter estimation is not performed. 

 In conclusion, the algorithm for the actual BMS needs to verify the estimation performance of the hybrid profile. In this paper, we focus on this part and verify the algorithm performance.

\subsection{Covariance (Estimator) wind-up problem}If there is insufficient excitation of the current input data and dynamic information sufficient to estimate the parameters of the battery is not input to the estimation algorithm, existing data is continuously forgotten. This phenomenon leads to exponential growth of the RLS covariance, which is another cause of the 'wind-up' problem.\unskip~\cite{342681:13001506} Therefore, the parameter estimation performance of the RLS is drastically reduced in the interval where the excitation is relatively insufficient among the driving patterns of the EV.  This disadvantage causes a reduction in the overall parameter estimation performance in the actual EV driving pattern that repeats the patterns of driving (dynamic pattern), charging and rest (static pattern).

\bgroup
\fixFloatSize{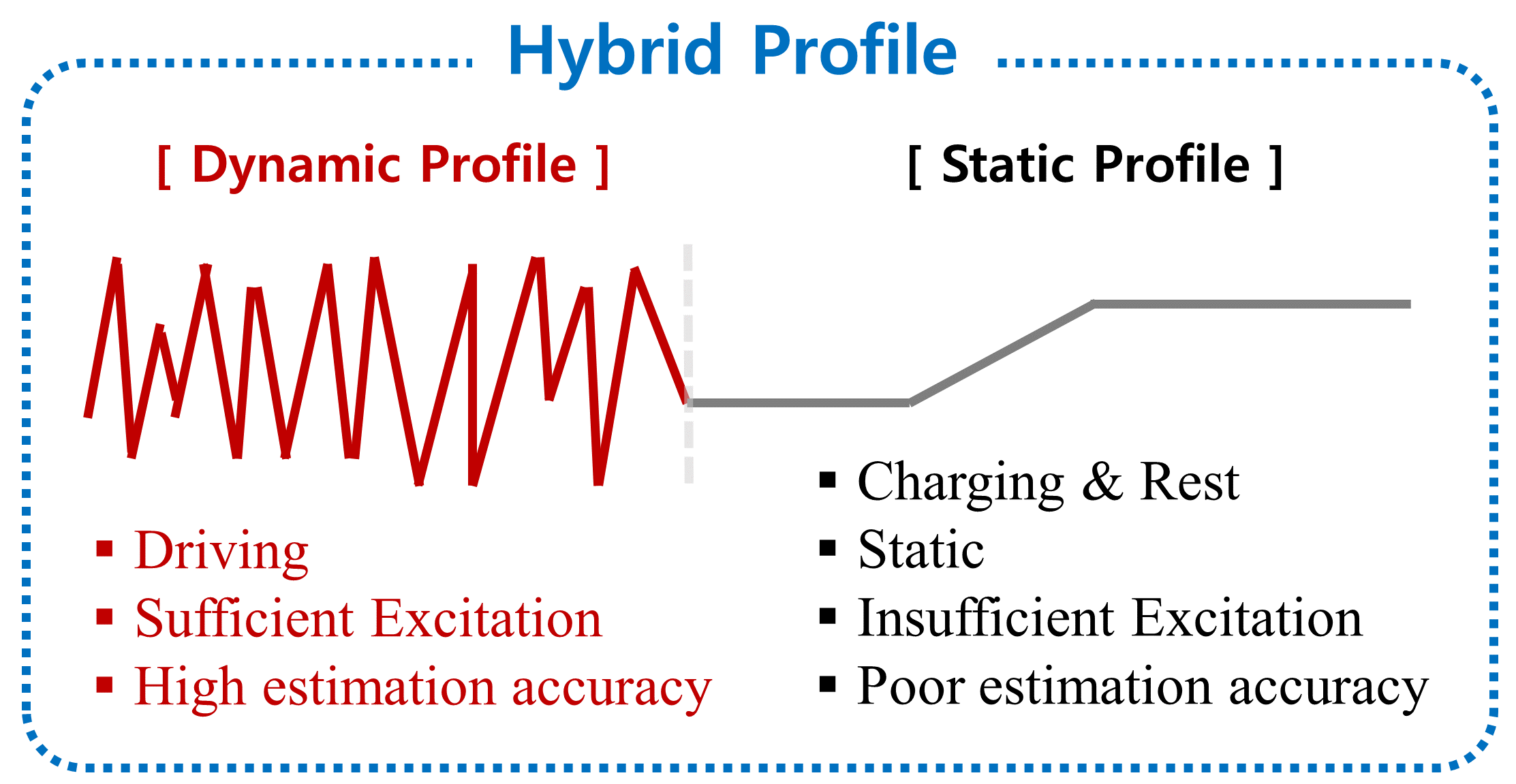}
\begin{figure}[!htbp]
\centering \makeatletter\IfFileExists{images/A8.png}{\includegraphics{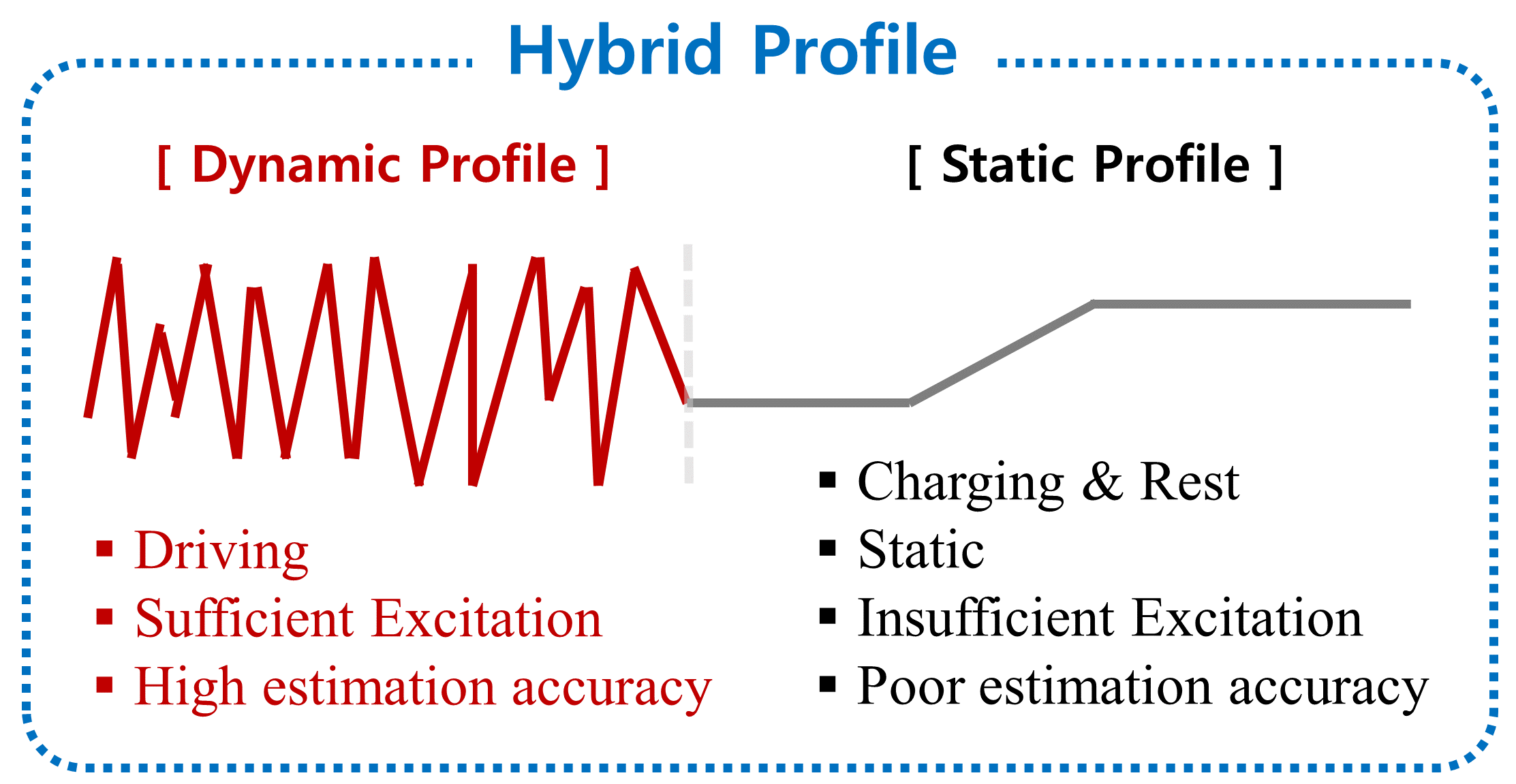}}{}
\makeatother 
\caption{{Hybrid profile}}
\label{f-02743f77313b}
\end{figure}
\egroup

\subsection{Condition number (CN) and Auto-tuning method} When estimating the battery parameters in the EV driving cycle using RLS, the accuracy of parameter estimation fluctuates greatly depending on the FV. For example, if a FV corresponding to a rapidly varying parameter has a small value (typically less than 0.8), the estimation accuracy of the parameter decreases. Therefore, setting this value is very important. At this time, the optimal value of FF greatly changes according to the driving pattern of the EV, the characteristics of the battery, and the change pattern of the corresponding parameter.

Previous studies using joint estimation used a forgetting factor as a fixed value. There are papers on how to set this value,\unskip~\cite{342681:15364208,342681:15364209} but they do not adequately reflect the characteristics of the system used. In actual commercial BMS, it is generally known to use the best result obtained by preprocessing that performs fine-tuning through several trials and errors on specific data. However, in an actual EV environment where driving patterns change every time and patterns cannot be predicted, this approach is not optimal.

 Therefore, in this paper, we applied an auto-tuning method that can automatically estimate the optimal FV in real time in real environment.\unskip~\cite{342681:16954759} Through this technique, the SOC/SOH estimation accuracy is not significantly affected by the value of the initial FV. It also has high estimation accuracy for various driving patterns. In this study, auto-tuning is performed for the first value which has the greatest influence on the estimation result among four FV to reduce the computation amount.

At this time, condition number (CN) was introduced as a measure for accurate parameter estimation.  CN is a measure to judge the sensitivity of the estimated parameter and has a lower value as the estimated parameter is robust to the change of the input.\unskip~\cite{342681:13001548} In this case, we can judge that the CN parameter has a more accurate RC parameter. \unskip~\cite{342681:13446884,342681:13446885} Based on this, in the auto-tuning algorithm, the CN of the parameter estimated by RLS is obtained, and the FFV is updated in the direction of decreasing CN. By applying the optimal FFV, SOC \& SOH estimation performance and estimation performance can be shown.

CN = $cond(A) $ represents the sensitivity of solution x according to the change of A, B in equation $Ax\;=\;B $. \mbox{}\protect\newline The basic definition of CN is :
\let\saveeqnno\theequation
\let\savefrac\frac
\def\dispfrac{\displaystyle\savefrac}
\begin{eqnarray}
\let\frac\dispfrac
\gdef\theequation{3.1}
\let\theHequation\theequation
\label{dfg-1ce2}
\begin{array}{@{}l}\frac { \frac { \left\ensuremath{\Vert} \Delta x \right\ensuremath{\Vert}  }{ \left\ensuremath{\Vert} x \right\ensuremath{\Vert}  }  }{ \frac { \left\ensuremath{\Vert} \Delta A \right\ensuremath{\Vert}  }{ \left\ensuremath{\Vert} A \right\ensuremath{\Vert}  } +\frac { \left\ensuremath{\Vert} \Delta B \right\ensuremath{\Vert}  }{ \left\ensuremath{\Vert} B \right\ensuremath{\Vert}  }  } \quad \le \quad cond(A)\end{array}
\end{eqnarray}
\global\let\theequation\saveeqnno
\addtocounter{equation}{-1}\ignorespaces 
The two matrices A and B used to estimate the parameters in the DFF-RLS are:
\let\saveeqnno\theequation
\let\savefrac\frac
\def\dispfrac{\displaystyle\savefrac}
\begin{eqnarray}
\let\frac\dispfrac
\gdef\theequation{3.2}
\let\theHequation\theequation
\label{dfg-51e2}
\begin{array}{@{}l}A\quad =\quad \sum _{ t=0 }^{ t }{ { \Lambda  }^{ t-i }{ \varphi  }_{ i }{ \varphi  }_{ i }^{ T } } \end{array}
\end{eqnarray}
\global\let\theequation\saveeqnno
\addtocounter{equation}{-1}\ignorespaces 

\let\saveeqnno\theequation
\let\savefrac\frac
\def\dispfrac{\displaystyle\savefrac}
\begin{eqnarray}
\let\frac\dispfrac
\gdef\theequation{3.3}
\let\theHequation\theequation
\label{dfg-0925}
\begin{array}{@{}l}B\quad=\quad\sum_{t=0}^{t}\Lambda^{t-i}\varphi_iy\end{array}
\end{eqnarray}
\global\let\theequation\saveeqnno
\addtocounter{equation}{-1}\ignorespaces 

\let\saveeqnno\theequation
\let\savefrac\frac
\def\dispfrac{\displaystyle\savefrac}
\begin{eqnarray}
\let\frac\dispfrac
\gdef\theequation{3.4}
\let\theHequation\theequation
\label{dfg-10af}
\begin{array}{@{}l}x = \theta\end{array}
\end{eqnarray}
\global\let\theequation\saveeqnno
\addtocounter{equation}{-1}\ignorespaces 
A and B are matrices of current and voltage, including errors due to measurement noise, disturbance, and round-off. Therefore, it is possible to determine the reliability of the parameter estimated by DFF-RLS through the CN. For example, when $cond(A) $is high, it is known that the estimated parameter x is sensitive to errors of A and B, so that x is a value far from the true value. Therefore, the reliability of x at this time can be judged to be low.

\bgroup
\fixFloatSize{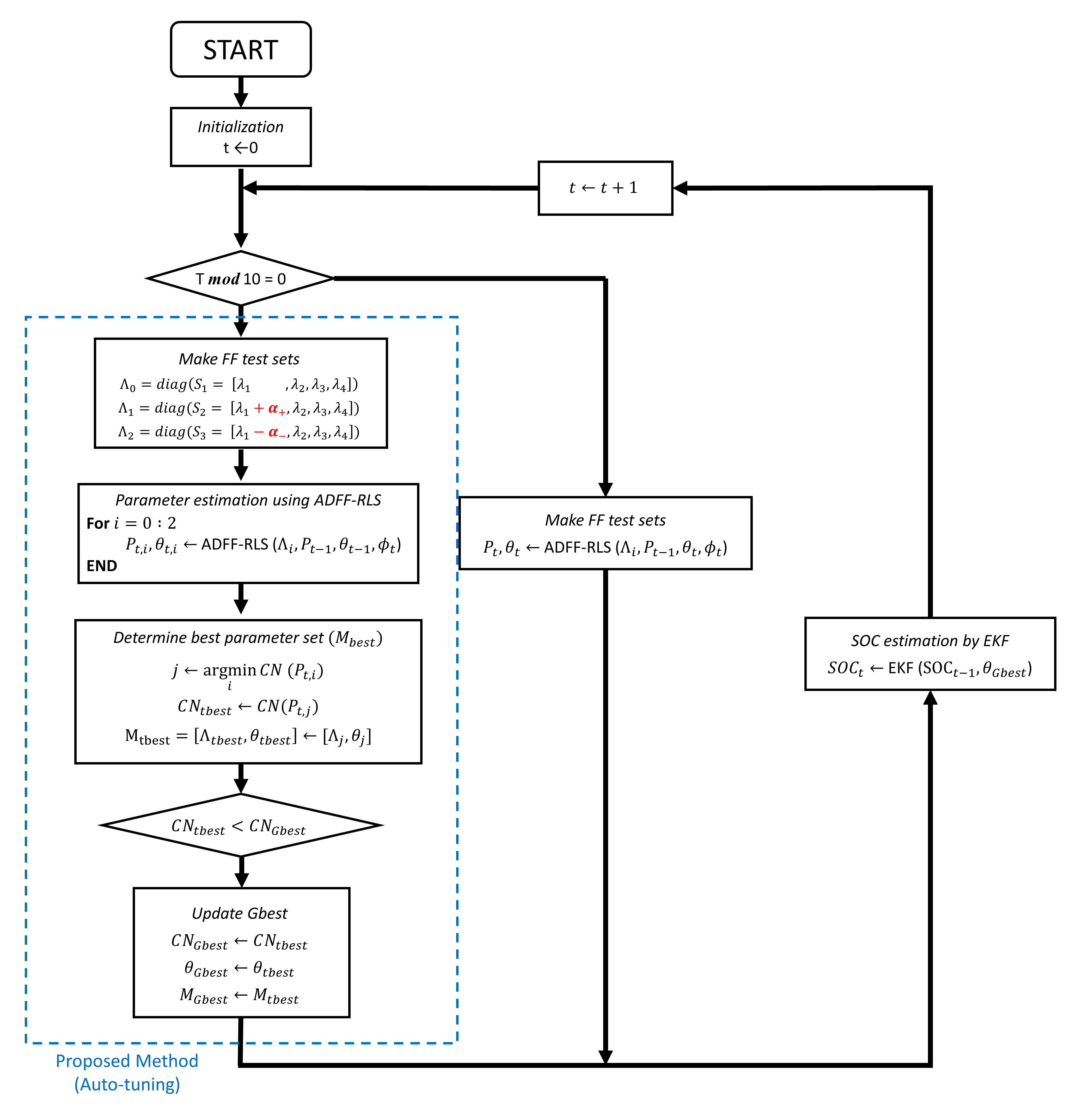}
\begin{figure*}[!htbp]
\centering \makeatletter\IfFileExists{images/A3.png}{\includegraphics{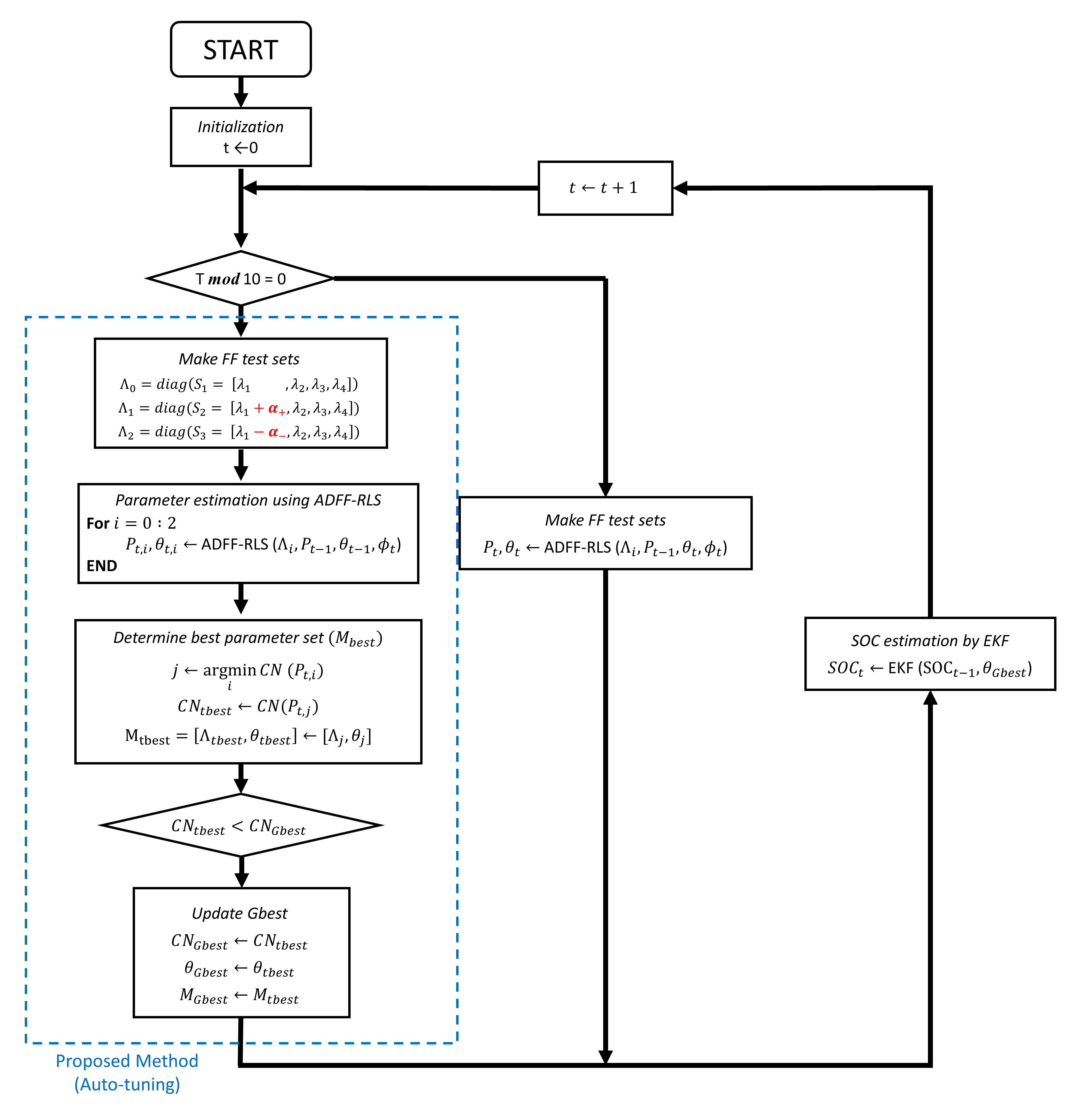}}{}
\makeatother 
\caption{{Algorithm Flowchart of auto-tuning method}}
\label{f-02ff69c89310}
\end{figure*}
\egroup

\subsection{Excitation-tag method} If there is insufficient excitation of the current input data and dynamic information sufficient to estimate the parameters of the battery is not input to the estimation algorithm, existing data is continuously forgotten. This phenomenon leads to exponential growth of the RLS covariance, which is another cause of the 'wind-up' problem. Therefore, the parameter estimation performance of the RLS is drastically reduced in the interval where the excitation is relatively insufficient among the driving patterns of the EV. This drawback becomes a problem in the actual EV driving pattern that repeats these patterns {\textemdash} Driving (dynamic pattern), Charging \& Rest (static pattern).

\bgroup
\fixFloatSize{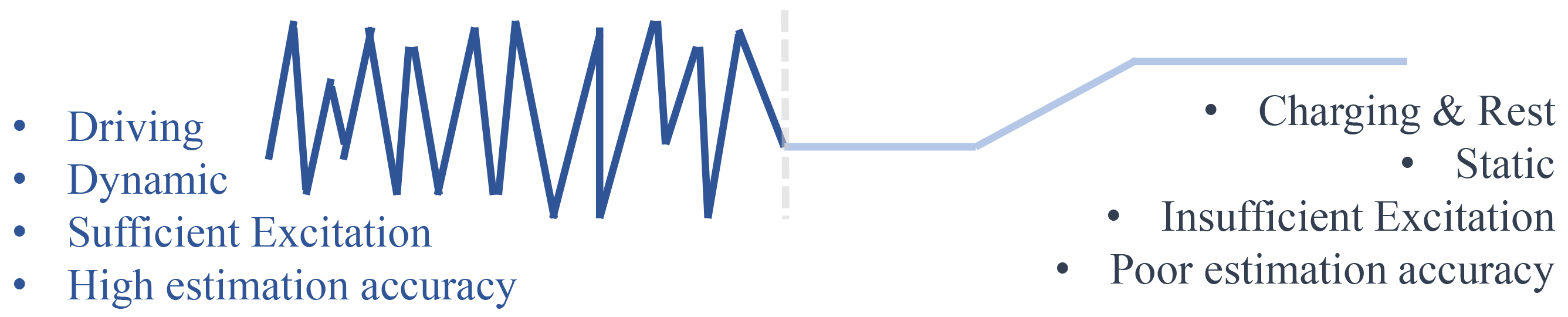}
\begin{figure}[!htbp]
\centering \makeatletter\IfFileExists{images/A7.png}{\includegraphics{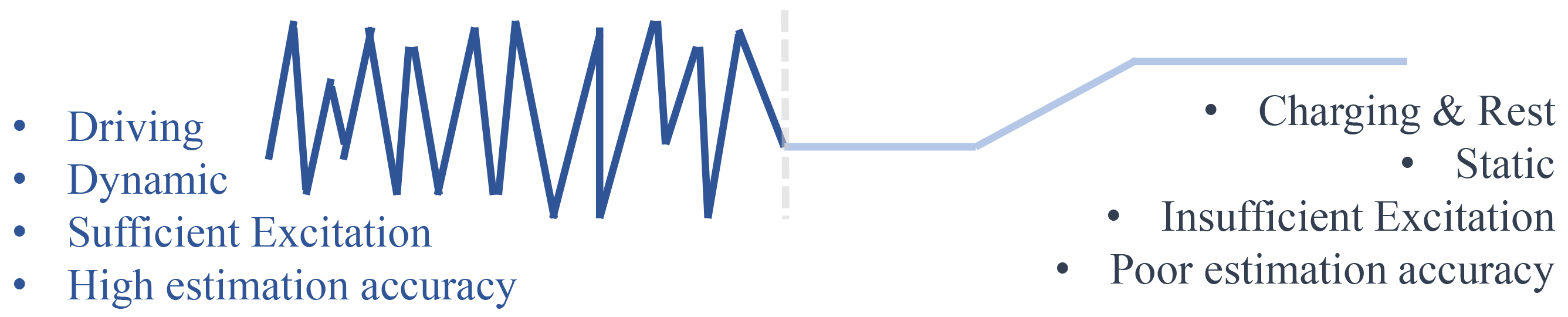}}{}
\makeatother 
\caption{{Ys}}
\label{f-283f11fdbeaf}
\end{figure}
\egroup
To solve this problem, a different technique can be applied to each section of the battery profile with different excitations, so that the information on the battery dynamics can be concentrated on the sections with sufficient excitation and the sections with insufficient excitation. To apply other estimation techniques to Based on this idea, we proposed the following solution-'Excitation Tag'. The concept is simple. The algorithm first determines whether excitation is sufficient or not enough, and then tags accordingly. (For example, enough sections are set to 1 and insufficient sections are set to 0.) The state estimation algorithm is then changed according to the tag.

 Depending on the excitation tag, the way the DFF-RLS and EKF sections work is different. When the excitation tag is 1, since the reliability of the ECM model parameter estimated by the RLS with the present input signal is high, the parameter is updated to the EKF. In this case, since the prediction value obtained through the model in the EKF is also highly reliable, the value of Q corresponding to the model prediction is lowered. Conversely, when the excitation tag is 0, the reliability of the parameter estimated through the RLS is low, so the parameter is not updated to the EKF, and the Q value at this time also increases. When this method is applied, the SOC/SOH estimation performance in the hybrid driving profile can be significantly improved.

\bgroup
\fixFloatSize{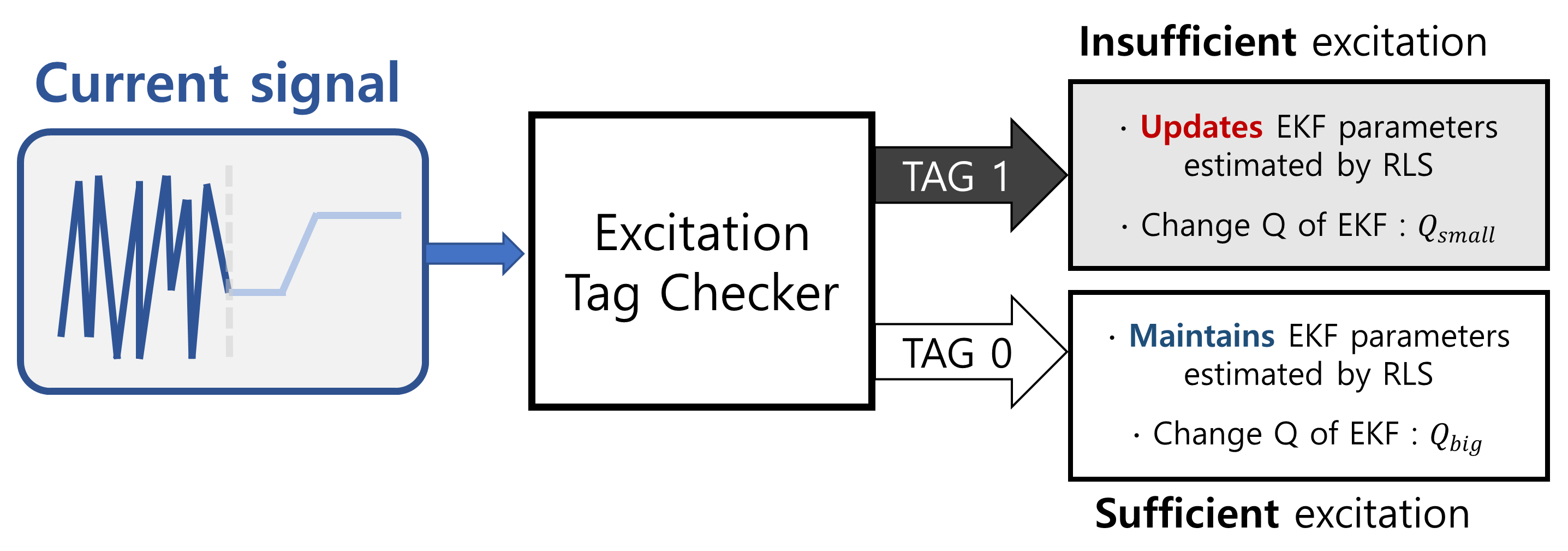}
\begin{figure}[!htbp]
\centering \makeatletter\IfFileExists{images/A4.png}{\includegraphics{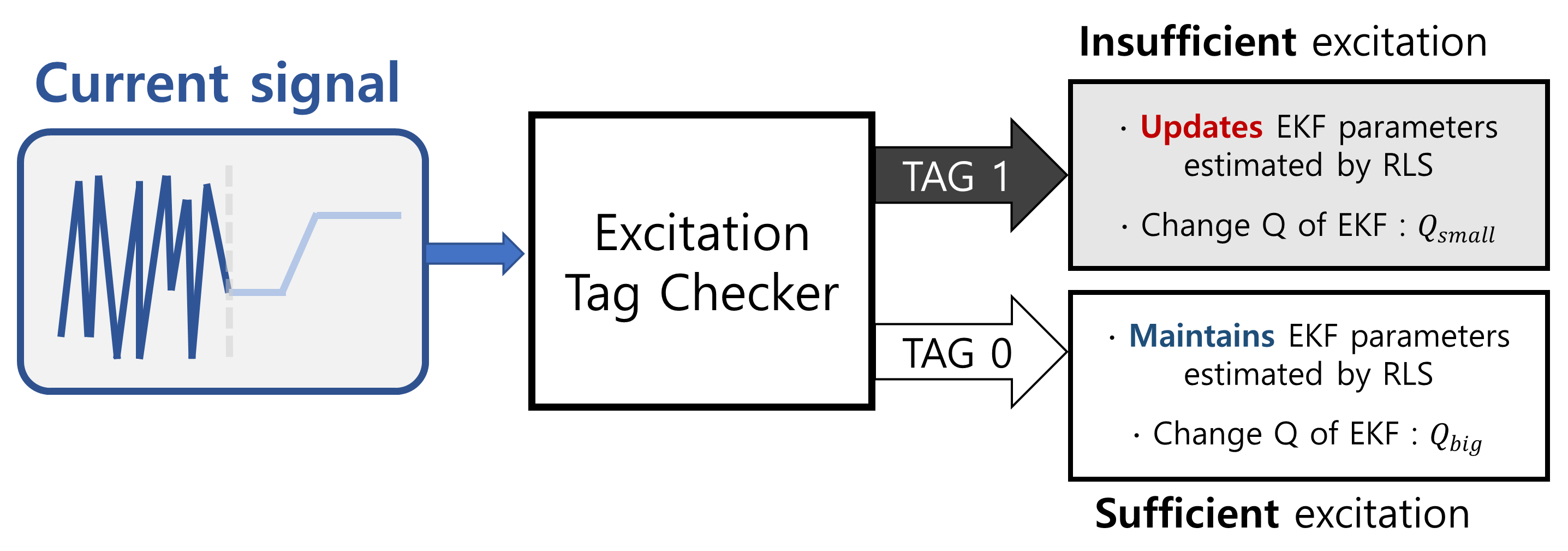}}{}
\makeatother 
\caption{{Excitation tag checker}}
\label{f-dc040d5c60b8}
\end{figure}
\egroup

\subsection{Adaptive DFF-RLS (ADFF-RLS)}We proposed Adaptive Diagonal Forgetting Factor Recursive Least Square (ADFF-RLS) which improved the existing DFF-RLS. The contribution of this proposal is as follows.

\bgroup
\fixFloatSize{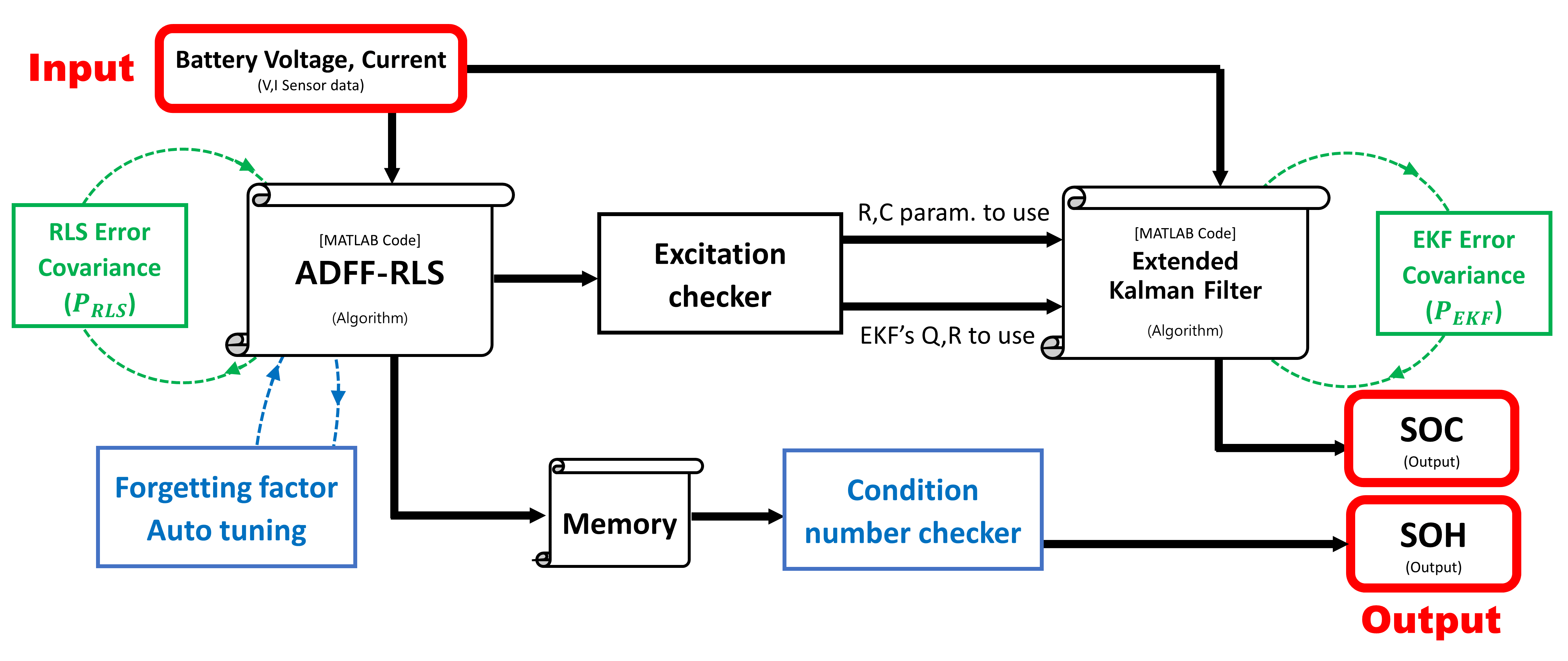}
\begin{figure*}[!htbp]
\centering \makeatletter\IfFileExists{images/A6.png}{\includegraphics{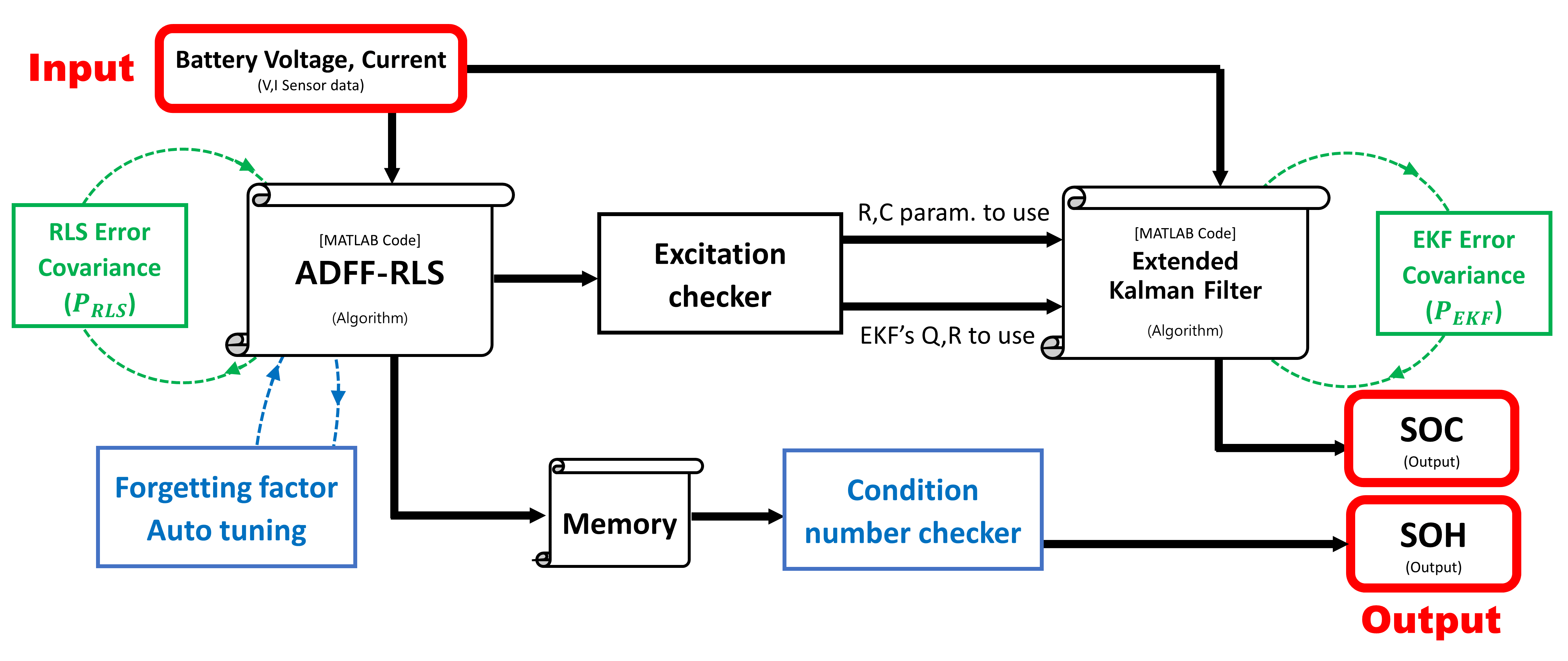}}{}
\makeatother 
\caption{{Flowchart of ADFF-RLS}}
\label{f-79bc95a8302b}
\end{figure*}
\egroup

  \begin{enumerate}
  \item \relax The covariance wind-up problem, a chronic problem of RLS, was solved by applying DFF-RLS.
  \item \relax Combining DFF-RLS and EKF, it was possible to obtain high SOC and SOH estimation performance compared to other algorithms (Single EKF, MFF-RLS \& EKF) while compensating for the shortcomings of each algorithm.
  \item \relax By applying the auto-tuning method of the forgetting factor of RLS through the condition number, it is possible to automatically and adaptively apply the optimal forgetting factor without fine-tuning the input data used for estimation.
  \item \relax Excitation tag that adaptively changes the tuning parameter of EKF according to the excitation of the input data is newly proposed to improve the estimation performance more than the existing EKF.
  \end{enumerate}
   To verify the performance of the proposed algorithm, the results were verified with the battery data used in the actual EV in the next section.
    
\section{Results}
This section shows a comparative study between the common SOC/SOH estimation method and the proposed ADFF-RLS method. Experimental data have been collected from a Li-ion (NMC) battery pack used in actual EV uses various data with different temperature from SOH.

\subsection{ Experimental test setup}Experimental data used Li-ion NMC 72Ah battery pack used in actual EV. The battery was tested with various current input profiles (Static and dynamic profile) and various SOH (Beginning of life) and EOL (End of life). The SOC/SOH was estimated using the voltage (V), current (I) and temperature (T) data obtained from this experiment.


\subsection{Error comparison result}
\begin{table}[!htbp]
\caption{{Driving profile with different SOH and noise} }
\label{tw-a0fda6458faf}
\def\arraystretch{1}
\ignorespaces 
\centering 
\begin{tabulary}{\linewidth}{p{\dimexpr.25\linewidth-2\tabcolsep}p{\dimexpr.3244\linewidth-2\tabcolsep}p{\dimexpr.1756\linewidth-2\tabcolsep}p{\dimexpr.25\linewidth-2\tabcolsep}}
\hline \cAlignHack Profile number & \cAlignHack Test profile & \cAlignHack Life status & \cAlignHack  Temperature\\
\hline 
\cAlignHack 3-1 &
  \cAlignHack Driving - GB/T &
  \cAlignHack BOL &
  \cAlignHack 30\ensuremath{^\circ}C\\
\cAlignHack 3-2 &
  \cAlignHack Driving - GB/T &
  \cAlignHack EOL &
  \cAlignHack 30\ensuremath{^\circ}C\\
\cAlignHack 3-3 &
  \cAlignHack Driving - US06 &
  \cAlignHack BOL &
  \cAlignHack 30\ensuremath{^\circ}C\\
\cAlignHack 3-4 &
  \cAlignHack Driving - GB/T+US06 &
  \cAlignHack EOL &
  \cAlignHack 30\ensuremath{^\circ}C\\
\hline 
\end{tabulary}\par 
\end{table}

\begin{table}[!htbp]
\caption{{Maximum and average absolute estimation error of SOC} }
\label{tw-8ac14cbc7db0}
\def\arraystretch{1}
\ignorespaces 
\centering 
\begin{tabulary}{\linewidth}{LLLLLLL}
\hline \cAlignHack \cellcolor[HTML]{B3B3B3}{Test profile} & \multicolumn{2}{p{\dimexpr(\mcWidth{2}+\mcWidth{3})}}{\cAlignHack \cellcolor[HTML]{B3B3B3}{Single EKF}} & \multicolumn{2}{p{\dimexpr(\mcWidth{4}+\mcWidth{5})}}{\cAlignHack \cellcolor[HTML]{B3B3B3}{MFF-RLS\&EKF}} & \multicolumn{2}{p{\dimexpr(\mcWidth{5}+\mcWidth{6})}}{\cAlignHack \cellcolor[HTML]{B3B3B3}{DFF-RLS\&EKF}}\\
\hline 
\cAlignHack \cellcolor[HTML]{E5E5E5}{Error [\%]} &
  \cAlignHack \cellcolor[HTML]{E5E5E5}{MAX} &
  \cAlignHack \cellcolor[HTML]{E5E5E5}{AVG} &
  \cAlignHack \cellcolor[HTML]{E5E5E5}{MAX} &
  \cAlignHack \cellcolor[HTML]{E5E5E5}{AVG} &
  \cAlignHack \cellcolor[HTML]{E5E5E5}{MAX} &
  \cAlignHack \cellcolor[HTML]{E5E5E5}{AVG}\\
\cAlignHack Driving - GB/T (BOL) &
  \cAlignHack 4.211 &
  \cAlignHack 1.805 &
  \cAlignHack 1.650 &
  \cAlignHack 1.315 &
  \cAlignHack 0.060 &
  \cAlignHack 0.048\\
\cAlignHack Driving - GB/T (EOL) &
  \cAlignHack 6.450 &
  \cAlignHack 2.763 &
  \cAlignHack 16.734 &
  \cAlignHack 2.603 &
  \cAlignHack 0.459 &
  \cAlignHack 0.144\\
\cAlignHack Driving - US06 (BOL) &
  \cAlignHack 6.649 &
  \cAlignHack 3.204 &
  \cAlignHack 7.648 &
  \cAlignHack 4.727 &
  \cAlignHack 0.644 &
  \cAlignHack 0.497\\
\cAlignHack Driving - GB/T+US06 (EOL) &
  \cAlignHack 5.355 &
  \cAlignHack 2.665 &
  \cAlignHack 1.552 &
  \cAlignHack 1.152 &
  \cAlignHack 0.017 &
  \cAlignHack 0.013\\
\hline 
\end{tabulary}\par 
\end{table}

\begin{table}[!htbp]
\caption{{Maximum and average absolute estimation error of estimated voltage (SOH error)} }
\label{tw-a174a019501a}
\def\arraystretch{1}
\ignorespaces 
\centering 
\begin{tabulary}{\linewidth}{LLLLLLL}
\hline \cAlignHack \cellcolor[HTML]{B3B3B3}{Test profile} & \multicolumn{2}{p{\dimexpr(\mcWidth{2}+\mcWidth{3})}}{\cAlignHack \cellcolor[HTML]{B3B3B3}{Single EKF}} & \multicolumn{2}{p{\dimexpr(\mcWidth{4}+\mcWidth{5})}}{\cAlignHack \cellcolor[HTML]{B3B3B3}{MFF-RLS\&EKF}} & \multicolumn{2}{p{\dimexpr(\mcWidth{5}+\mcWidth{6})}}{\cAlignHack \cellcolor[HTML]{B3B3B3}{DFF-RLS\&EKF}}\\
\hline 
\cAlignHack \cellcolor[HTML]{E5E5E5}{Error [mV]} &
  \cAlignHack \cellcolor[HTML]{E5E5E5}{MAX} &
  \cAlignHack \cellcolor[HTML]{E5E5E5}{AVG} &
  \cAlignHack \cellcolor[HTML]{E5E5E5}{MAX} &
  \cAlignHack \cellcolor[HTML]{E5E5E5}{AVG} &
  \cAlignHack \cellcolor[HTML]{E5E5E5}{MAX} &
  \cAlignHack \cellcolor[HTML]{E5E5E5}{AVG}\\
\cAlignHack Driving - GB/T (BOL) &
  \cAlignHack 60.600 &
  \cAlignHack 4.620 &
  \cAlignHack 72.913 &
  \cAlignHack 8.792 &
  \cAlignHack 27.854 &
  \cAlignHack 0.337\\
\cAlignHack Driving - GB/T (EOL) &
  \cAlignHack 56.220 &
  \cAlignHack 9.523 &
  \cAlignHack 2916.000 &
  \cAlignHack 18.775 &
  \cAlignHack 49.428 &
  \cAlignHack 4.815\\
\cAlignHack Driving - US06 (BOL) &
  \cAlignHack 110.300 &
  \cAlignHack 7.968 &
  \cAlignHack 126.390 &
  \cAlignHack 22.616 &
  \cAlignHack 208.130 &
  \cAlignHack 10.602\\
\cAlignHack Driving - GB/T+US06 (EOL) &
  \cAlignHack 41.130 &
  \cAlignHack 8.299 &
  \cAlignHack 42.354 &
  \cAlignHack 15.714 &
  \cAlignHack 37.584 &
  \cAlignHack 1.774\\
\hline 
\end{tabulary}\par 
\end{table}
 Standardized car driving test profile GB / T and US06, and EKF alone (Single EKF) for battery test data tested with a mixture of two profiles, previously proposed MFF-RLS + EKF, and proposed in this study Algorithm (ADFF-RLS + EKF) was applied and the results were compared and verified. From the overall data, we can see that the proposed algorithm has higher estimation accuracy.
    
\section{Conclusions}
In this paper, we proposed a real-time joint estimation (DMFF-RLS \& EKF) algorithm using excitation tag and verified by experiment. The contribution of this proposal is as follows.  

Through the joint estimation of DFF-RLS and EKF, it can adapt to changing environments in EV real-time driving environment and obtain high SOC / SOH estimation performance.

  \begin{enumerate}
  \item \relax Condition number was introduced to numerically represent the accuracy of the estimated parameters and auto-tuning was performed based on the condition number. Therefore, high estimation accuracy can be obtained without fine-tuning of tuning parameters, and adaptive response can be achieved through automatic tuning even if the driving pattern changes.
  \item \relax Through the above proposals, high estimation accuracy can be obtained by improving 'covariance wind-up problem' which is a disadvantage of RLS-based estimation method.
  \end{enumerate}
  In order to evaluate the accuracy of the SOC/SOH estimation result of this algorithm, we applied the hybrid driving pattern cell data of real EV's with two SOH (BOL \& EOL) to the proposed method, and confirmed the SOC estimation results for each data. The same data were compared with several conventional algorithms. The proposed algorithm was more accurate than the other algorithms. These results were common to all battery life types. Therefore, it is confirmed that the proposed algorithm has higher SOC/SOH estimation accuracy than the existing algorithm in the environment where the traveling pattern changes in real time. This result is expected to provide more accurate information about the driver. Therefore, the driver can manipulate the EV in a more economical way and obtain an improved driving experience.


\section*{Acknowledgements}


%




\bibliography{article}
\bibliographystyle{IEEEtran}

%




\vskip -2\baselineskip plus -1fil%
\begin{IEEEbiographynophoto}{Kwangrae Kim}
    received the B.S. degree inelectrical engineering from Konkuk University, Seoul, South Korea, in 2015, and the M.S. degree in electrical engineering from the Pohang Uni-versity of Science and Technology (POSTECH),Pohang, South Korea, in 2018, where he is cur-rently pursuing the Ph.D. degree in electrical engineering.  His  research  interests  include ithium-ion battery state estimation, battery man-agement systems, and optimization applications in battery field.
\end{IEEEbiographynophoto}

%




\vskip -2\baselineskip plus -1fil%
\begin{IEEEbiographynophoto}{Minho Kim}
    received the B.S. degree in mechan-ical engineering from Korea University, Seoul,South Korea, in 2015. He is currently pursuingthe Ph.D. degree with the Department of CreativeIT Engineering, Pohang University of Science andTechnology (POSTECH), Pohang, South Korea.His main research interests include energy storagesystems and machine learning technology
\end{IEEEbiographynophoto}

%




\vskip -2\baselineskip plus -1fil%
\begin{IEEEbiographynophoto}{Suwon Kang}
    received the B.S. degree in electrical engineering from Kyungpook national university, Daegu, South Korea, in 2015, and the M.S. degree in electrical engineering from the Pohang University of Science and Technology (POSTECH), Pohang, South Korea, in 2017. He is currently working in the LG chem. His research interests include lithium-ion battery state estimation, functional safety engineering in automotive battery field.
\end{IEEEbiographynophoto}

%




\vskip -2\baselineskip plus -1fil%
\begin{IEEEbiographynophoto}{Jungwook Yu}
    received the B.S. degree ininformation management and technology from Syracuse University, NY, USA, in 2012. He iscurrently  pursuing  the  Ph.D.  degree  withthe  Department  of  Creative  IT  Engineering, Pohang University of Science and Technology(POSTECH), Pohang, South Korea. His research interests include battery management systems, parameter estimation, and neural networks
\end{IEEEbiographynophoto}

%




\vskip -2\baselineskip plus -1fil%
\begin{IEEEbiographynophoto}{Jungsoo Kim}
    was born in Seoul, South Korea. He received the B.S. degree in mechanical engineering from Korea University, Seoul, SouthKorea, in 2013. He is currently pursuing the Ph.D.degree (M.S{\textendash}Ph.D. integrated program) with the Department of Creative IT Engineering (CiTE), Pohang University of Science and Technology(POSTECH), Pohang, South Korea. His research interests include li-ion battery state estimation, li-ion battery parameter identification, battery management systems, and their applications to electric vehicles and smartgrids.
\end{IEEEbiographynophoto}

%




\vskip -2\baselineskip plus -1fil%
\begin{IEEEbiographynophoto}{Huiyong Chun}
    was born in Seoul, SouthKorea, in 1994. He received the B.S. degree in electrical engineering from the Pohang University of Science and Technology (POSTECH), Pohang,South Korea, in 2017, where he is currently pursuing the Ph.D. degree with the Department of Creative IT Engineering. His research inter-ests include lithium-ion battery state and parameter estimation, battery management systems, and machine learning applications in lithium-ionbattery
\end{IEEEbiographynophoto}

\end{document}